# MARS15 CODE DEVELOPMENTS DRIVEN BY THE INTENSITY FRONTIER NEEDS[*][†]

N.V. Mokhov[a#], P. Aarnio[b], Yu.I. Eidelman[a†], K.K. Gudima[c], A.Yu. Konobeev[d],
V.S. Pronskikh[a§], I.L. Rakhno[a], S.I. Striganov[a], I.S. Tropin[a]

[a]*Fermi National Accelerator Laboratory, Batavia IL 60510-5011, USA*
[b]*Department of Applied Physics, Aalto University, FI-00076, Finland*
[c]*Institute of Applied Physics, Academy of Sciences of Moldova*
[d]*Institut für Reaktorsicherheit, Forschungszentrum Karlsruhe GmbH,
Herrmann-von-Helmholtz-Platz 1, Eggenstein-Leopoldshafen 76344 Germany*

## Abstract

The MARS15(2012) is the latest version of a multi-purpose Monte-Carlo code developed since 1974 for detailed simulation of hadronic and electromagnetic cascades in an arbitrary 3-D geometry of shielding, accelerator, detector and spacecraft components with energy ranging from a fraction of an electronvolt to 100 TeV. Driven by needs of the intensity frontier projects with their Megawatt beams, e.g., ESS, FAIR and Project X, the code has been recently substantially improved and extended. These include inclusive and exclusive particle event generators in the 0.7 to 12 GeV energy range, proton inelastic interaction modeling below 20 MeV, implementation of the EGS5 code for electromagnetic shower simulation at energies from 1 keV to 20 MeV, stopping power description in compound materials, new module for DPA calculations for neutrons from a fraction of eV to 20-150 MeV, user-friendly DeTra-based method to calculate nuclide inventories, and new ROOT-based geometry.

[*]Work supported by Fermi Research Alliance, LLC under contract No. DE-AC02-07CH11359 with the U.S. Department of Energy.
[†]Presented paper at the 12th International Conference on Radiation Shielding, September 2-7, 2012, Nara, Japan
[#]mokhov@fnal.gov

# MARS15 code developments driven by the intensity frontier needs

Nikolai Mokhov[a*], Pertti Aarnio[b], Yury Eidelman[a†], Konstantin Gudima[c], Alexander Konobeev[d],
Vitaly Pronskikh[a§], Igor Rakhno[a], Sergei Striganov[a], Igor Tropin[a]

[a]*Fermi National Accelerator Laboratory, Batavia IL 60510-5011, USA*
[b]*Department of Applied Physics, Aalto University, FI-00076, Finland*
[c]*Institute of Applied Physics, Academy of Sciences of Moldova*
[d]*Institut für Reaktorsicherheit, Forschungszentrum Karlsruhe GmbH,
Herrmann-von-Helmholtz-Platz 1, Eggenstein-Leopoldshafen 76344 Germany*

The MARS15(2012) is the latest version of a multi-purpose Monte-Carlo code developed since 1974 for detailed simulation of hadronic and electromagnetic cascades in an arbitrary 3-D geometry of shielding, accelerator, detector and spacecraft components with energy ranging from a fraction of an electronvolt to 100 TeV. Driven by needs of the intensity frontier projects with their Megawatt beams, e.g., ESS, FAIR and Project X, the code has been recently substantially improved and extended. These include inclusive and exclusive particle event generators in the 0.7 to 12 GeV energy range, proton inelastic interaction modeling below 20 MeV, implementation of the EGS5 code for electromagnetic shower simulation at energies from 1 keV to 20 MeV, stopping power description in compound materials, new module for DPA calculations for neutrons from a fraction of eV to 20-150 MeV, user-friendly DeTra-based method to calculate nuclide inventories, and new ROOT-based geometry.

## 1. Introduction

MARS is a multi-purpose Monte-Carlo code developed since 1974 for detailed simulation of hadronic and electromagnetic cascades in an arbitrary 3-D geometry of shielding, accelerator, detector and spacecraft components with energy ranging from a fraction of an electronvolt to 100 TeV [1]. Driven by needs of the Intensity Frontier (IF) projects with their Megawatt beams, the code has been recently substantially improved and extended. The developments crucial for design of the IF accelerators, beamlines, targets and precision experiments are: particle production models in medium-energy nuclear interactions (including the near-threshold kaon production), an accurate description of low-energy electromagnetic showers and electromagnetic interactions of particles and heavy ions down to 1 keV/A in compounds (energy deposition, radiation damage and backgrounds), radiation damage models as well as nuclide production, decay and transmutation for inventory and residual dose estimations. These new developments in MARS15(2012) are described in this paper along with two new general-purpose modules which enhance the code capabilities: the ROOT-based geometry module and extended Graphical User Interface and visual editor.

## 2. Inclusive and exclusive modeling and event generators

Most of processes in MARS15, such as electromagnetic showers, hadron-nucleus interactions, decays of unstable particles, emission of synchrotron photons, photohadron production and muon pair production, can be treated exclusively (analogously), inclusively (with corresponding statistical weights), or in a mixed mode. The choice of method is left for the user to decide - via the input settings – what is the most appropriate and computationally efficient for the considered physics case.

Charged pion production is described with high precision by the two-source model in a few GeV energy range. The double-differential cross section can be fitted by the following nine-parameter formula with $\chi^2/n < 2$:

$$E\frac{d^3\sigma}{dp^3} = p_1(1+ p_7 \cos\theta) \exp\left[-\frac{T(1-p_2\cos\theta)}{p_3}\right]$$
$$+ \frac{p_9(1+p_8\cos\theta)}{1+p_4 \exp\left[-T(1-p_6\cos\theta)/p_5\right]}.$$

---

*Corresponding author. Email: mokhov@fnal.gov
†On leave from Budker Institute of Nuclear Physics SB RAS, Novosibirsk 630090, Russia.
§On leave from Joint Institute for Nuclear Research, Dubna, Moscow region 141980, Russia.

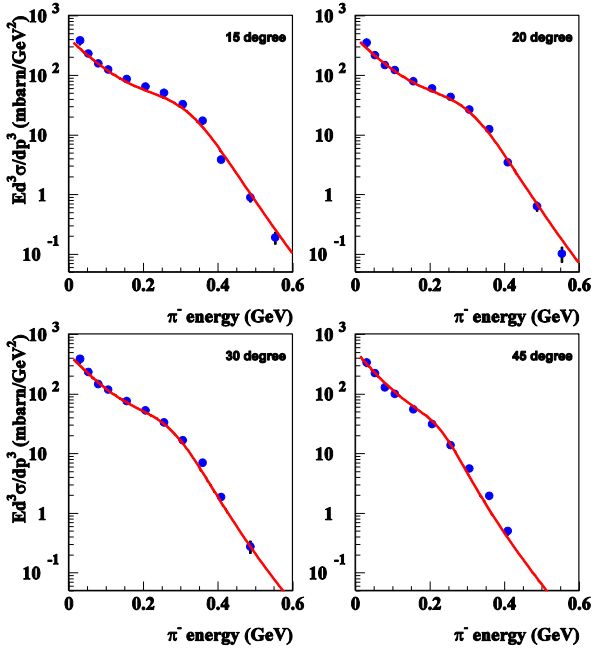

Figure 1. Negative pion production in proton-lead collisions at 730 MeV.

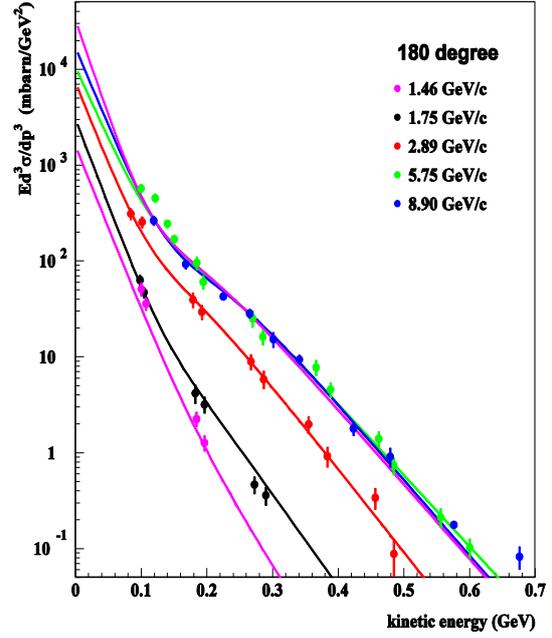

Figure 2. Negative pion production in proton-lead collisions.

Fig. 1 compares results of this approximation with the LANL measurements [2]. Experimental data of the HARP collaboration [3] covers angles from 20° to 123°. This data could be also successfully fitted by the two-source model with $\chi^2/n \sim 1$ for any HARP momentum (3, 5, 8, 12 GeV/c). If other measurements at large angles are included into fitting procedure, quality of the fit becomes slightly worse ($\chi^2/n \sim 2$), but still is acceptable. Fig. 2 presents comparison of the two-source model with data at different proton momenta. It is seen that the proposed parameterization could be applied in broad angular and momentum ranges. It is known that intra-nuclear cascade models have problems with description of the HARP data [3]. The two-source model can be used as an inclusive event generator until LAQGSM [4] will improve description of the low-energy pion production.

The LAQGSM module in MARS15 is based on the quark-gluon string model above 10 GeV and intranuclear cascade, pre-equilibrium and evaporation models at lower energies [4]. It was modified to improve its performance in the crucial for the Intensity Frontier energy region of 0.7 to 12 GeV. The newest developments include: new and better approximations for elementary total, elastic, and inelastic cross sections for NN and πN interactions; several channels have been implemented for an explicit description: N+N→N+N+mπ, π+N→N+mπ (m<5), B+B→B+Y+K, π+B→Y+K, Kbar+B→Y+π, and K+Kbar, N+Nbar pair production; combination of the phase space and isobar models and experimental data; γA reactions extended down to GDR and below; and an arbitrary light nuclear projectile (*e.g.*, d) and nuclear target (*e.g.*, He). Fig. 3 shows comparison of calculated kaon spectra with experimental data [5].

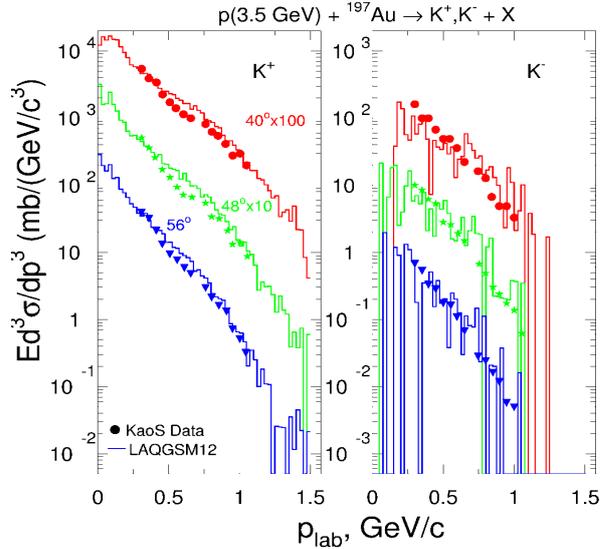

Figure 3. Kaon production cross section for 3.5-GeV protons on a gold nucleus.

ALICE2011 [6] is the nuclear model code based on hybrid model of precompound decay, Weisskopf-Ewing evaporation and Bohr-Wheeler fission models. It was improved and converted to an event generator for nucleon, photon and heavy-ion nuclear reactions at E~1 MeV to 20-30 MeV matching CEM and LAQGSM at E above 20-30 MeV in MARS15. Fig. 4 shows that an experimental charge-exchange reaction cross section on $^{65}$Cu is reproduced well practically down to the threshold energy at 2.168 MeV.

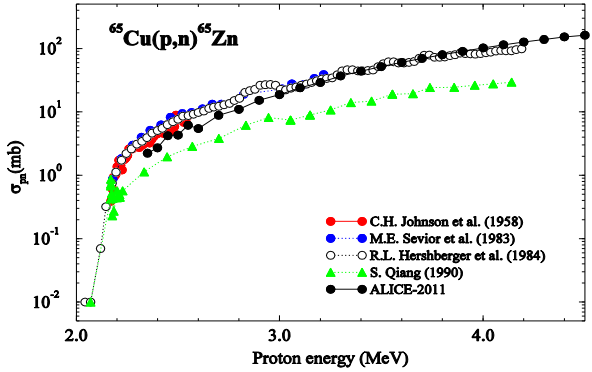

Figure 4. Comparison between measured and calculated (ALICE-2011) (p,n) reaction cross sections on $^{65}$Cu.

*3. Electromagnetic interactions*

Stopping power of ions in compounds is usually described according to the Bragg's rule. At low energies and for low-Z materials the difference between measured and predicted dE/dx can be as large as 20%. The "cores-and-bonds" method (CAB), developed by G. Both *et al.* [7], was implemented in MARS15 taking into account chemical bonds fitted to experiment for various compounds at 1 keV to 3 MeV [8]. Fig. 5 shows accuracy of the two methods compared to experiment. At higher energies, the Sternheimer and Peierls density correction algorithm for compounds is employed [9].

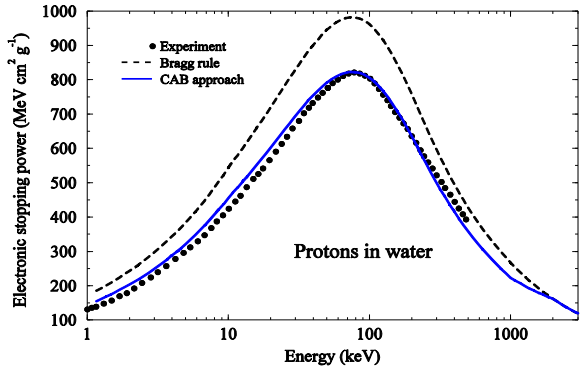

Figure 5. A comparison between experimental data [10] and theoretical predictions for proton dE/dx in water.

Inclusive, exclusive and hybrid modeling of electromagnetic showers at all energies is now controlled in a user-friendly way globally or for specified materials. The EGS5 code [11] has been implemented in MARS15 for precise modeling of electromagnetic showers in the 1 keV to 20 MeV energy range (see Fig. 6) globally or in specified materials. It is crucial for accurate description of transition effects in fine accelerator and detector structures, background studies and medical applications. MARS takes care of EGS5 initialization, shares geometry, magnetic fields and estimation modules. Users have a full control over EGS5 simulation via PEGS input files, initially generated by MARS with some default values. Comparisons of energy deposition in copper calculated with different codes are shown in Fig. 7.

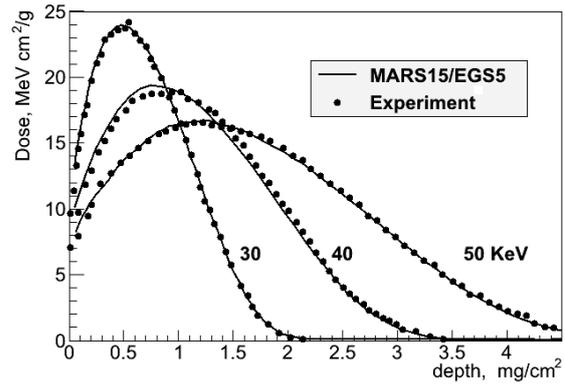

Figure 6. Measured [12] and simulated distributions of absorbed dose for an electron beam impinging normally on a surface of semi-infinite silicon.

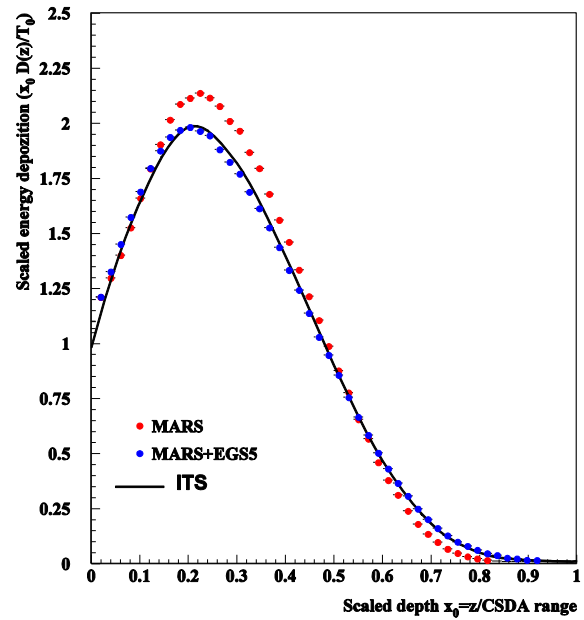

Figure 7. Calculated energy depositions due to 5-MeV electron beam in Cu. The ITS results are taken from Ref. [13].

*4. Nuclide inventory*

The DeTra code [14] integrated in MARS15 provides now a user-friendly way for a 3-step nuclide decay and transmutation analysis to calculate nuclide inventory: (1) a standard MARS run for nuclide production and stopping rates in materials specified with interface files NUCLIDES generated; stopping rates are crucial for fine structures; (2) built-in DeTra is called to solve the Bateman equations governing the decay and transmutation of nuclides using transmutation trajectory analysis; (3) the output files of step 2 are processed to order specific activities and production rates. The output of the third step can be directly used to calculate corresponding residual dose rates [15]. Fig. 8 shows comparisons between measured and predicted residual activity in a copper target. The MARS15 performance is quite impressive.

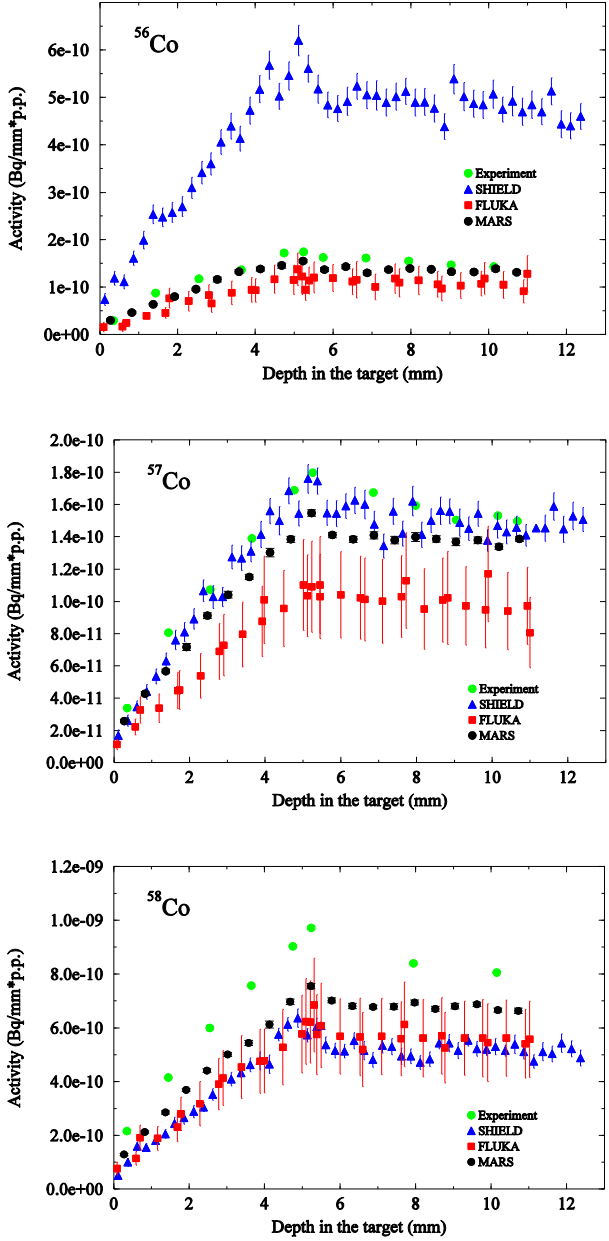

Figure 8. Comparisons between measured [16] and calculated longitudinal distributions of specific residual activity of cobalt isotopes generated and stopped in a copper target irradiated with a 500 MeV/u uranium beam of 11 mm in diameter. The target length was twice the range of the uranium ions and transverse target size was 50 mm. It was assembled using copper disks and the activation foils were inserted between the disks. Statistical errors for MARS15 correspond to 2.5 million histories; statistical errors for SHIELD and FLUKA are taken from Ref. [16].

## 5. Displacement-per-atom (DPA)

A new model was developed for neutrons from $10^{-5}$ eV to 20-150 MeV using the NJOY99+ENDF-VII database for 393 nuclides. Results of calculations with NRT model are corrected using experimental defect production efficiency η, where η is a ratio of a number of single interstitial atom vacancy pairs (Frenkel pairs) produced in a material to the number of defects calculated using NRT model. The values of η have been measured experimentally [17] for many important materials for a number of neutron fluxes in reactor energy range.

In this approach, the number of displacements, corresponding to the NRT model, is calculated with the use of defect production cross sections (Fig. 9) based on ENDFB-VII library. Following that, cross sections are corrected for η (Fig. 10), scaling them to the experimental ones.

The MARS15 model of displacements, produced by elastic Coulomb scattering of charged particles, was improved by modification of the screening parameter. Range of applicability of the model has been significantly increased, especially for low-energy heavy particles.

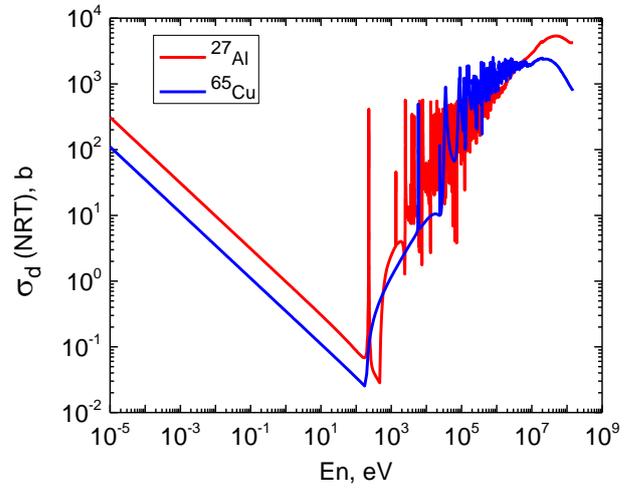

Figure 9. NRT neutron defect production cross-sections for $^{27}$Al and $^{65}$Cu calculated using NJOY99 code.

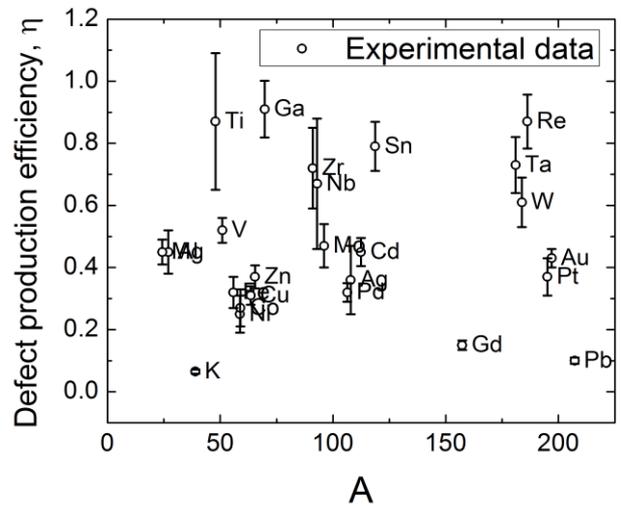

Figure 10. Measured [16] defect production efficiency η for a number of materials.

## 6. ROOT geometry mode

The powerful ROOT geometry and visualization options [18] have been implemented into MARS15. Geometry models created for MARS15 can now be used with other Monte Carlo codes (*e.g.*, Geant4), and one can use the ROOT models created for Geant4 with MARS15. ROOT provides a large set of geometrical elements (primitives) along with a possibility to produce composite shapes and assemblies as well as 3D visualization. Examples of ROOT models built from scratch and imported to MARS15 are shown in Figs. 11 and 12, respectively.

information on the elements including magnetic field presented in a selected optics file, the beam line builder will generate a corresponding 3D geometry model. An example of such a model is shown in Fig. 13. For scoring information on radiation fields inside the beam line elements it is now possible to define histograms in a reference system where longitudinal axis follows axes of elements along the beam line, whereas transverse coordinates correspond to the ones for the local reference system of the beam line element. The so-called XYS HBOOK histograms can be defined in a corresponding XYZHIS.INP file.

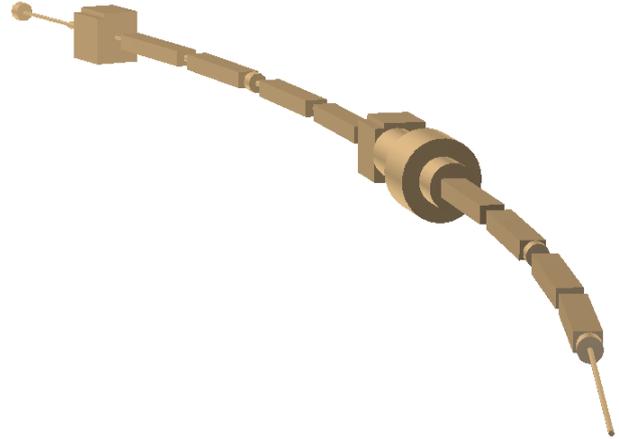

Figure 13. A 3D view for geometry of the Fermilab Booster section created by means of the ROOT-based beam line builder.

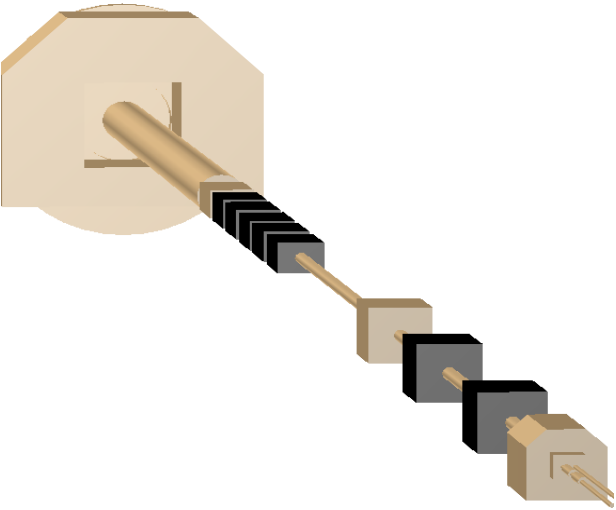

Figure 11. A ROOT model for a straight section of the LHC IR5 used in MARS15 for simulation of machine related backgrounds in the CMS detector.

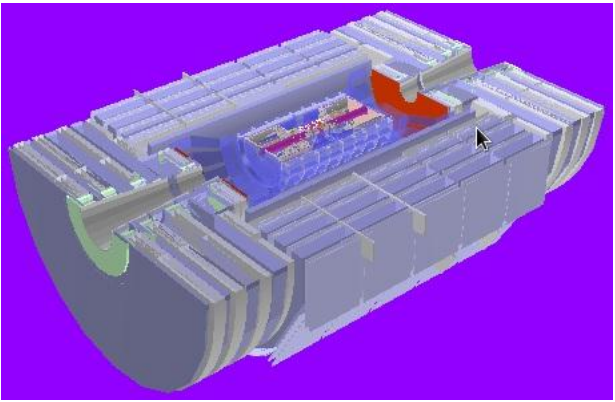

Figure 12. A 3D cut-away view of the CMS detector imported to MARS15 from a ROOT source tree.

The MARS beam line builder [19] was redesigned for using with the ROOT geometry packages. Usually, a beam line consists of elements of several pre-defined types: magnets (dipoles, quadrupoles etc.), correctors and collimators. A user has to describe how to build 3D geometry models for elements of each type. Using the

## 7. Graphical User Interface (GUI)

The MARS-GUI-Slice was further improved. Its major new features include the following: (i) a possibility to plot two sets of particle's tracks for the selected energy ranges; (ii) managing different objects (like labels, tick labels, user's labels *etc.*) in a GUI window; (iii) improved data output (quantitatively and qualitatively) in the relevant information windows like "Particles", "Point Info" and "Track Info" *etc.*; (iv) added ability to launch the code paw++ (management of histograms) and subroutine MARS2BML (to transform coordinates from the global MARS frame to a beam line frame).

A new visual editor was developed for the following files: MARS.INP, MATER.INP and GEOM.INP. It allows now interactive editing of the input data and checking its self-consistency and coherence, which greatly reduces the possibility of running with a wrong input (see Fig. 14). The performed syntax verification provides consistency of newly entered cards with the existing ones. In the "manual" editing mode, the same syntax verification is performed for all the cards being edited. A feature was implemented into the editor that allows one to take a look at all the three files MARS.INP, MATER.INP and GEOM.INP simultaneously. At the same time, instead of these default file names one can enter arbitrary names, and the editor will open files with the entered names assuming that the files have corresponding structures. Such a feature is useful when a user has, *e.g.*, different files with geometry description for the same system. When writing to the files, the editor supports an agreement on various default data in the cards, and the default data will be shown in appropriate fields during the editing session. Each card includes a description of the data it contains. The description is either a part of the corresponding section from the MARS user's guide [1], or completely coincides with it.

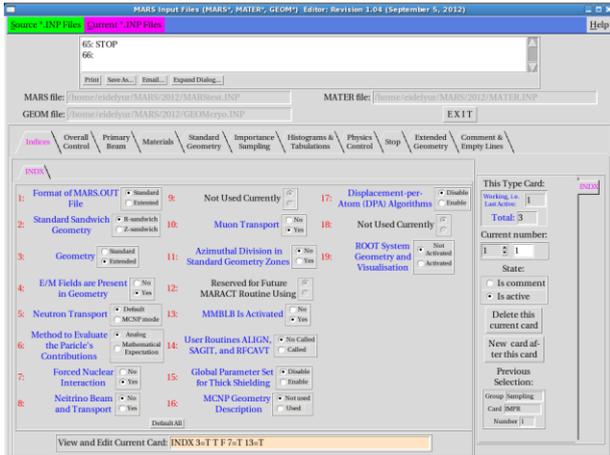

Figure 14. An example of an editing session with the new visual editor.

**Acknowledgements**

Work supported by Fermi Research Alliance, LLC, under contract DE-AC02-07CH11359 with the U.S. Department of Energy.